\documentstyle[prl,aps,epsf,psfig,cite]{revtex}
\input epsf.sty
\begin{document}

\draft
\title{ Deformation of the Fermi surface in the extended Hubbard model }
\author{B. Valenzuela $^1$ and M. A. H. Vozmediano $^2$ }
\address{
        $^1$Unidad Asociada ICMM-UC3M,
	Instituto de Ciencia de Materiales,
        Consejo Superior de
	Investigaciones Cient{\'\i}ficas.
        Cantoblanco, 28049 Madrid. Spain.
\\
        $^2$Unidad Asociada ICMM-UC3M,
	Departamento de Matem\'aticas,
	Universidad Carlos III de Madrid,
	Avda. de la Universidad 30, 28911 Legan\'es, Madrid, Spain.}

\date{\today}
\maketitle

\vspace{1cm}
\begin{abstract}

The deformation of the Fermi surface induced by Coulomb interactions 
is investigated in the $t-t'$--Hubbard
model. The interplay of  the local $U$ and extended $V$  
interactions is analyzed. 
It is found that  exchange interactions $V$ enhance small 
anisotropies producing deformations of the Fermi surface which break the point 
group symmetry of the square lattice
at the Van Hove filling .  This Pomeranchuck instability competes with ferromagnetism
and is suppressed at a critical value $U(V)$. The 
interaction $V$  renormalizes the $t'$ parameter 
to smaller values  what favors nesting. It also
induces changes on the topology of hole of the  Fermi surface which can go
from hole to electron-like
what may explain recent ARPES experiments. 

\end{abstract}

The shape of the Fermi surface  plays a crucial role in determining the
instabilities of interacting Fermi systems. The importance of Fermi surface 
deformations by interaction in the non-spherical case was already recognized in 
the early papers on the foundation of the Fermi liquid concept \cite{KL1,KL2}, 
and remains a central question both under a fundamental point of view as well 
as for the phenomenological implications.

The  physics of the cuprates and the recent ARPES experiments has renewed 
interest on the anisotropic Fermi surfaces. Main questions are how close 
is the meassured Fermi surface from the non--interacting?, how much can the 
Fermi surface be deformed before it breaks down driving
the system to a broken symmetry  phase?, does the deformation always  
respect the point group symmetry of the lattice? 

In this paper we address the former questions in a very simplified model 
of the cuprates. We study the Fermi surface deformation in the $t-t'$--Hubbard 
model on the squared two--dimensional lattice in the presence of week to 
moderate Coulomb interactions. We include an on-site coupling $U$ 
to take account of the magnetic interactions and 
an extended exchange interaction $V$ which can be as large as $t$ \cite{auer}
and is widely used recently to account for
the stripe feature and the possible phase separated phase in the cuprates.
It will be seen that the $V$ interaction plays 
a crucial role when the shape
of the Fermi surface does not have special features favoring magnetic instabilities.

The deformation of the Fermi surface in the Hubbard and $t-J$ models has been studied 
recently \cite{HM1,morita,himeda,nojiri} with a variety of methods. In most of these 
approaches attention is centered on the influence of magnetic interactions so the exchange  
interaction $V$ is not taken into account. Our results show that, even at the 
mean field level, this interaction has a strong influence on the Fermi 
surface renormalization.

In a very recent paper \cite{HM4} it was argued that the Fermi surface of the
$t-t'$-- Hubbard model may undergo a deformation breaking the point symmetry
of the underlying lattice (Pomeranchuck instability) driven by the renormalization
of the forward channel. A  deformation of this type seems to be observed in recent
ARPES experiments \cite{asensio}. Ferromagnetic instabilities which would directly compete
with the Pomeranchuck instability were not taken into account. 
We investigate the conditions for
the formation of Pomeranchuck instabilities and the competition with ferromagnetism.
We argue that the instability is a fine tuning effect which only takes place at the Van Hove
filling and is suppressed by ferromagnetism.

Our analysis provides also interesting results concerning the renormalization of 
the $t'$ parameter which is shown to  decrease with $V$.

The hamiltonian that we shall use is
\begin{equation}
H=-\sum_{i,j,\sigma} t_{ij} c^+_{i,\sigma}
c_{j,\sigma}
\;+\;U\sum_i n_{i,\uparrow} n_{i,\downarrow}\;+V\sum_{<ij>,ss'}n_{is}n_{js'}
 \;\;\;,
\label{ham}
\end{equation}
where $t_{ij}=t$ for nearest neighbors, $t_{ij}=t'$ for next nearest neighbors, and
it is zero otherwise, and where we have included an extended Coulomb interaction $V$.

The non-interacting band structure of (\ref{ham}) in the squared lattice is
\begin{equation}
\varepsilon^0({\bf k})=
-2t\cos k_x-2t\cos k_y+4t'\cos k_x\cos k_y 
\label{disp}
\end{equation}
what originates the constant energy contours shown in fig. 1.
\begin{figure}
\begin{center}
\mbox{\psfig{file=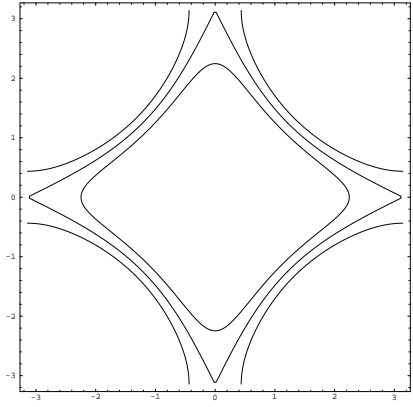,width=.3\textwidth}}
\end{center}
\caption{Fermi surface of the free system for t=1, t'=0.3
and various densities around the Van Hove filling.} 
\label{freefs}
\end{figure}
The shape of the interacting Fermi surface is encoded in the two--point 
Green's function of the interacting system
\begin{equation}
G(\omega,{\bf k})=\frac{1}
{\omega-(\varepsilon^0({\bf k})-\mu)-\Sigma(\omega,{\bf k})} \;\;,
\end{equation}
where $\Sigma(\omega,{\bf k})$ is the fermion self--energy.
The Fermi surface of the interacting system  is defined by 
the equation \cite{KL2}
\begin{equation}
\mu-\varepsilon^0({\bf k})-\Re\Sigma(\omega,{\bf k})=0\;\;,
\label{mu}
\end{equation}
where $\Re \Sigma$ stands for the real part of the electron self--energy.
A mean field decoupling of the interacting term in (\ref{ham}) gives rise to 
an exchange interaction 
\begin{equation}
V({\bf k}_x,{\bf k}_y)=
2V\sum_{k'_x,k'_y}[\cos (k_x-k'_x)+ \cos (k_y-k'_y)]n({\bf k'}) \;\;\;,
\label{V}
\end{equation}
where $n({\bf k})$ is the occupation of the site ${\bf k}$ in momentum space.
The ${\bf k}$ dependence of the interacion $V$ in (\ref{V}) is the key 
point of the results. 

\begin{figure}
\begin{center}
\mbox{\psfig{file=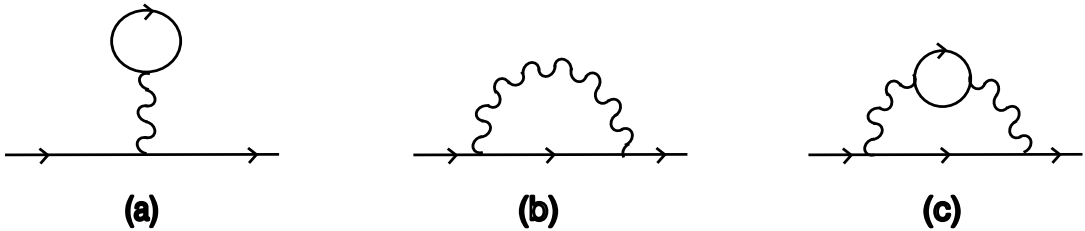,width=.8\textwidth}}
\end{center}
\caption{Feynman diagrams contributing to the electron self--energy at
lowest order in perturbation theory.} 
\label{graphs}
\end{figure}

The diagrams contributing to the electron self--energy in perturbation 
theory are shown in fig. \ref{graphs}.

Local interactions $U$ do not change the shape of the Fermi surface at the
one loop level. Diagrams in fig. 2a) and fig. 2b) give rise to a constant
that is absorbed in the chemical potential so as to keep the system at
constant density. The first contribution to the
Fermi surface deformation in the local case comes from the diagram in fig. 
2c) and has been computed in the literature \cite{HM1,nojiri} and found to be very small.
The interaction $V$ contributes to the fermion self--energy at the one loop level
through diagram in fig. 2b).

We compute numerically the contribution of diagram in fig. 2b) to the
Fermi surface deformation in eq. (\ref{mu}) by solving self-consistently the 
equation
\begin{eqnarray}
\varepsilon({\bf k}) &= & \varepsilon_{kin}({\bf k}) +\varepsilon_{exch}({\bf k})-\mu 
\nonumber \\
& = & -2t\cos k_x-2t\cos k_y+4t'\cos k_x\cos k_y - 
2V\sum_{k'_x,k'_y}[\cos (k_x-k'_x)+ \cos (k_y-k'_y)]n({\bf k'})  -\mu \;\;\;,
\label{nospin}
\end{eqnarray}
where $\mu$ is adjusted so that the total number of electrons remains constant and spin
interactions are not taken into account.

Ferromagnetism is included by starting with two different Fermi surfaces 
for electrons with spin up and down and solving the equations
\begin{eqnarray}
\varepsilon_\uparrow ({\bf k})& = & \varepsilon_\uparrow^0 ({\bf k})
+\varepsilon_\uparrow^{exch} ({\bf k})+U\frac{n_\downarrow}{N}-\mu \nonumber \\
\varepsilon_\downarrow ({\bf k})& = & \varepsilon_\downarrow^0 ({\bf k})
+\varepsilon_\downarrow^{exch} ({\bf k})+U\frac{n_\uparrow}{N}-\mu \;\;,
\label{ferro}
\end{eqnarray}
where the total number of electrons $N=n_\uparrow +n_\downarrow$ remains constant.

From eqs. (\ref{nospin}) and 
(\ref{ferro}), it is clear that in the absence of feromagnetism,
i.e starting with two identical Fermi surfaces in (\ref{ferro}),
the interaction $U$ contributes to the total energy with a global constant
that is absorbed in  the chemical potential. The deformations induced by $V$ are
in general similar to the ones obtained in the literature when computing diagram
in fig. 2c) with a local $U$ interaction. The reason is that the electron  
susceptibility inserted in the loop  acts as a ${\bf k}$--dependent
effective interation. The global sign is also the same (attractive).
In the second order perturbation theory because of the closed fermion
loop in the diagram, and in $V$ due to the exchange. The present analysis
has the relevance of acting already at the one loop level.  

Our results can be summarized as follows. The deformation
of the Fermi surface depends on the  filling. 
The general effect of $V$
is to smoothen out anisotropies in the curvature of the Fermi surface. This
has a number of interesting consequences. In
particular for closed surfaces (electron--like), the inflection points present
with a finite $t'$ which were related in \cite{us} to  pairing instabilities 
with an extended-s order parameter, dissapear at a value of $V\sim 1$. In the context
of ref. \cite{us}, the presence of $V$ favors a d-wave pairing. This
result seems at odds with the claims in \cite{himeda} which find spontaneous
generation of $t'$ in a Monte Carlo analysis of the $t-J$ model. We think that this
is due to the sign of the interaction. The effect of a positive $V$ is the reversed
one: it reinforces the curvature of the Fermi surface leading to 
bigger values of $t'$. 

The general agreement of our computation with other very different approaches 
\cite{morita} suggests that  there will be a general tendency towards nesting in
the $t-t'$--Hubbard model with an extended attractive interaction of any origin.

A related topic  of experimental relevance 
\cite{exp1} concerns the possible change 
in Fermi surface topology induced by interactions. The presence of a $t'$
parameter breaking the electron--hole symmetry of the Hubbard model is crucial in this
analysis.  We find no changes in the Fermi surface topology  
with $t'=0$ in agreement with \cite{HM1}. In the presence of finite $t'$ and
$V$, the Fermi surface does undergo changes in topology from open (hole--like) to 
closed  (electron--like) when the initial Fermi energy lies close to -- and slightly over --
the Van Hove filling.  This effect can explain recent messurements of electron--like surfaces
in BISCO \cite{exp1,exp2}. The reversed change would be observed by reversing the
sign of $V$.

It is interesting to note that  the interacting Fermi surface lies at the level of
the Van Hove singularities for a variety of initial fillings 
in the hole--like  regime.
The Van Hove filling is never reached if starting with an overdoped situation 
(closed Fermi surface). This last result is
very insensitive to the value of $t'$ and mimics the result about the
pinning of the Fermi surface obtained in \cite{nos}. 

We have searched for Pomeranchuck instabilities by investigating the response
of the system to  very small perturbations of the free Fermi surface. The result is 
extremely sensitive to the value of the chemical potential of the final system.
The effect of the  interactions is, in general, to restore the lattice 
symmetry for generic values of the final $\mu$.  Even in the
case of very strong initial perturbation (of the order of a 4 per cent), the final
Fermi surface is four-fold symmetric. 
The exception arises when the chemical potential
of the interacting system coincides with the Van Hove singularity . 
We have checked that, in the absence of ferromagnetism,
the Van Hove filling is unstable towards Pomeranchuck
deformations. This result agrees with the general considerations exposed in
\cite{and} that the Fermi level of electron systems tend to avoid 
peaks in the density of states.

\begin{figure}
\begin{center}
\mbox{\psfig{file=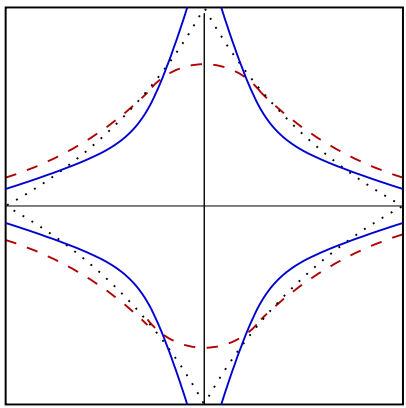,width=.3\textwidth}}
\end{center}
\caption{Deformation of the Fermi surface induced by  $V$ and by a small anisotropy in
the free band. Solid line: free FS; dotted line: final FS in the absence of 
anisotropies; dashed line: deformed FS.} 
\label{vh}
\end{figure}

Fig. \ref{vh} shows the deformation caused by $V=2$ in a paramagnetic situation 
when a point is removed by hand in the initial Fermi surface of a  61 x 61 lattice
(corresponding to an anisotropy of less than $10^{-3}$). 

The solid line represents the free Fermi surface, the dotted line is the interacting Fermi
surface in the absence of the perturbation, and the dashed line shows the final Fermi surface
obtained after perturbing the free Fermi surface. Similar results are obtained with
different values of V(n). This result is independent of $t'$ and occurs also at 
$t'=0$. It is nevertheless extremely sensitive to the value of the final
chemical potential which has to be fine tuned by selecting the initial
number of  particles. 

Magnetic interactions do substantially change the former picture.  
We have centered our attention on the effect of ferromagnetism as  the predominant 
competing interaction in the forward channel \cite{us2}. Our results show that
ferromagnetism is enhanced by $U$ and $t'$ as expected, but it is disfavored  by $V$. 
This can be understood on the light of our previous analysis showing that $V$
renormalizes the $t'$ parameter downwards. For small to  moderate values of $U$,
the free ferromagnetic system evolves to an interacting  paramagnetic state. 
There is, however, a critical value of $U(V)$ where the final state corresponds to a 
fully polarized ferromagnetic system.  This supression of
ferromagnetism by  exchange interactions might explain the absence of 
ferromagnetic phases in real
systems which are otherwise well described by the $t-t'$--Hubbard model as the one in
\cite{us2}. 

\begin{figure}
\begin{center}
\mbox{\psfig{file=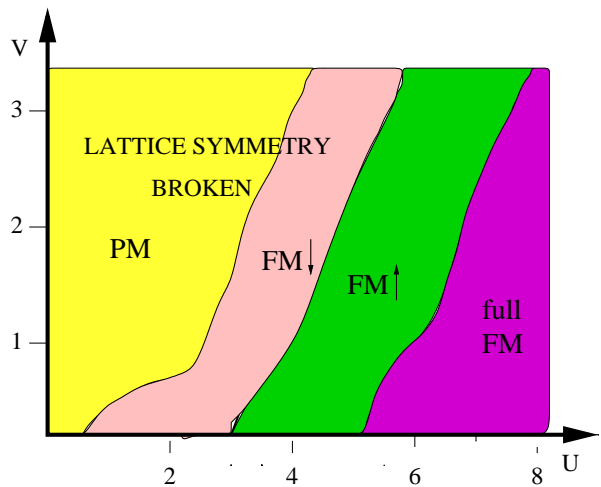,width=.45\textwidth}}
\end{center}
\caption{Schematic evolution of ferromagnetism with the interactions. PM: paramagnetic phase,
FM$\uparrow$ (FM$\downarrow$) : magnetic polarization of the interacting system smaller 
(bigger) than that of the free system; FM: fully polarized final state.} 
\label{phased}
\end{figure}

The Pomeranchuck instabilities described above compete with ferromagnetism. 
A schematic phase diagram showing the  evolution of the ferromagnetic behavior of 
the system as a function of $U$ and $V$ is despicted in  fig. \ref{phased}.

Pomeranchuck instabilities are found systematically under the appropriate conditions
in the paramagnetic zone of the phase diagram. By increasing $U$ and keeping $V$ constant,
they enter into the
phase denoted by FM$\downarrow$ in fig. \ref{phased} changing it into a paramagnetic phase.  
Finally there is a critical value of
$U$ for which ferromagnetism prevails and the deformation dissapears.

In conclusion we have shown that  exchange interactions  renormalize $t'$ to smaller 
values. An inmediate consequence
of this is that, in our search for microscopic models of the cuprates, one
should start the computations with a bigger value of $t'$ to account for a
phenomenologically fitted value of $t'$.  The effect is doping--dependent and should be
stronger in the underdoped situation where screening is less effective. In this regime
it will increase the nesting of the Fermi surface and enhance antiferromagnetic instabilities.
The renormalization of the $t'$ parameter may also explain the 
absence of ferromagnetic phases in some real systems described by the $t-t'$--Hubbard
model, and recent meassurements of electron--like Fermi surfaces in some cuprates.

Pomeranchuck instabilities are found as
the response of the system to small impurities when the Fermi surface lies at the Van 
Hove filling and can be regarded as an instability of the Van Hove level.  
They compete with ferromagnetism and
dissapear at a critical value U(V).

Although the results presented in this paper  are mean field results , we believe that the 
main features  will survive the  test with more accurate methods  and that exchange 
interactions should be included -- or checked to be negligible -- in 
Hubbard related studies.

\vspace{0.5cm}
\noindent
{\bf Acknowledgements.}
We are very grateful to F. Guinea for valuable discussions. 
Financial support from MEC (Spain) through grant
PB96/0875 and CAM (Madrid) through grant 07/0045/98 is 
acknowledged.

\end{document}